\PassOptionsToPackage{unicode}{hyperref}
\PassOptionsToPackage{hyphens}{url}
\documentclass[
  sigconf,
  screen,
  nonacm]{acmart}
\usepackage{xcolor}
\ifdefined\Bbbk
  \let\Bbbk\relax
\fi
\usepackage{amsmath,amssymb}
\makeatletter
\@ifundefined{KOMAClassName}{
  \IfFileExists{parskip.sty}{%
    \usepackage{parskip}
  }{
    \setlength{\parindent}{0pt}
    \setlength{\parskip}{6pt plus 2pt minus 1pt}}
}{
  \KOMAoptions{parskip=half}}
\makeatother
\usepackage{color}
\usepackage{fancyvrb}

\DefineVerbatimEnvironment{Highlighting}{Verbatim}{commandchars=\\\{\}}
\newenvironment{Shaded}{}{}

\newcommand{\AttributeTok}[1]{\textcolor[rgb]{0.49,0.56,0.16}{#1}}

\newcommand{\ControlFlowTok}[1]{\textcolor[rgb]{0.00,0.44,0.13}{\textbf{#1}}}
\newcommand{\DataTypeTok}[1]{\textcolor[rgb]{0.56,0.13,0.00}{#1}}

\newcommand{\ErrorTok}[1]{\textcolor[rgb]{1.00,0.00,0.00}{\textbf{#1}}}
\newcommand{\ExtensionTok}[1]{#1}

\newcommand{\FunctionTok}[1]{\textcolor[rgb]{0.02,0.16,0.49}{#1}}

\newcommand{\KeywordTok}[1]{\textcolor[rgb]{0.00,0.44,0.13}{\textbf{#1}}}
\newcommand{\NormalTok}[1]{#1}
\newcommand{\OperatorTok}[1]{\textcolor[rgb]{0.40,0.40,0.40}{#1}}

\newcommand{\StringTok}[1]{\textcolor[rgb]{0.25,0.44,0.63}{#1}}

\NewDocumentCommand\citeproctext{}{}

\makeatletter
 \let\@cite@ofmt\@firstofone
 \def\@biblabel#1{}
 \def\@cite#1#2{{#1\if@tempswa , #2\fi}}
\makeatother
\newlength{\cslhangindent}
\newlength{\csllabelwidth}
\newenvironment{CSLReferences}[2] 
 {\begin{list}{}{%
  \setlength{\itemindent}{0pt}
  \setlength{\leftmargin}{0pt}
  \setlength{\parsep}{0pt}
  \ifodd #1
   \setlength{\leftmargin}{\cslhangindent}
   \setlength{\itemindent}{-1\cslhangindent}
  \fi
  \setlength{\itemsep}{#2\baselineskip}}}
 {\end{list}}
\usepackage{calc}

\providecommand{\tightlist}{%
  \setlength{\itemsep}{0pt}\setlength{\parskip}{0pt}}
\usepackage{float}
\usepackage{placeins}
\renewcommand\footnotetextcopyrightpermission[1]{}
\makeatletter\let\@copyrightspace\relax\makeatother
\usepackage{fvextra}
\RecustomVerbatimEnvironment{Highlighting}{Verbatim}{commandchars=\\\{\},breaklines=true,breakanywhere=true,fontsize=\small}
\fvset{breaklines=true,breakanywhere=true}
\acmConference[arXiv 2026]{arXiv Preprint}{March 2026}{Bloomington, IN, USA}
\acmYear{2026}\copyrightyear{2026}
\acmBooktitle{arXiv Preprint (March 2026)}
\acmPrice{0.00}
\acmDOI{}
\acmISBN{}
\usepackage{bookmark}
\IfFileExists{xurl.sty}{\usepackage{xurl}}{} 
\hypersetup{
  pdftitle={Quine: Realizing LLM Agents as Native POSIX Processes},
  hidelinks,
  pdfcreator={LaTeX via pandoc}}

\title{Quine: Realizing LLM Agents as Native POSIX Processes}
\author{Hao Ke\\
Independent Researcher, Bloomington, IN, USA\\
i@kehao.me}
\date{March 2026}

\begin{document}
\begin{abstract}
Current LLM agent frameworks often implement isolation, scheduling, and
communication at the application layer, even though these mechanisms are
already provided by mature operating systems. Instead of introducing
another application-layer orchestrator, I present Quine, a runtime
architecture and reference implementation that realizes LLM agents as
native POSIX processes. The mapping is explicit: identity is PID,
interface is standard streams and exit status, state is memory,
environment variables, and filesystem, and lifecycle is fork/exec/exit.
A single executable implements this model by recursively spawning fresh
instances of itself. By grounding the agent abstraction in the OS
process model, Quine inherits isolation, composition, and resource
control directly from the kernel, while naturally supporting recursive
delegation, self-renewal through exec, and shell-native composition. The
design also exposes where the POSIX process model stops: processes
provide a robust substrate for execution, but not a complete runtime
model for cognition. In particular, the analysis points toward two
immediate extensions beyond process semantics: task-relative worlds and
revisable time. A reference implementation of Quine is publicly
available at https://github.com/kehao95/quine.
\end{abstract}
\maketitle

\section{Introduction}\label{introduction}

\begin{quote}
``Write programs that do one thing and do it well. Write programs to
work together. Write programs to handle text streams, because that is a
universal interface.'' --- Doug McIlroy (McIlroy 1978)
\end{quote}

\subsection{The Problem}\label{the-problem}

The question of how to structure LLM agents is often asked at the
framework level. This paper asks it one layer lower: what if the runtime
substrate were not an application framework, but the operating system
itself?

The rapid development of LLM agents has led to a proliferation of
frameworks designed to manage their lifecycle, memory, and
communication. Systems such as LangChain (LangChain, Inc. 2023), AutoGen
(Wu et al. 2023), and CrewAI (CrewAI, Inc. 2024) have successfully
democratized agent development by providing high-level abstractions for
these tasks. However, a structural pattern has emerged: these systems
primarily expose agent abstractions at the application layer, even when
the underlying execution still relies on OS services.

This pattern adds complexity that grows with system scale. By managing
agents as objects within a single application process, frameworks
frequently reconstruct mechanisms the operating system already provides:

\begin{itemize}
\tightlist
\item
  \textbf{Fault isolation} is simulated through exception handlers and
  try-catch blocks, rather than enforced by hardware-backed address
  space separation.
\item
  \textbf{Context switching} between agents requires application-level
  scheduling logic, rather than delegating to the kernel's mature
  scheduler.
\item
  \textbf{Message passing} is mediated by in-process queues or
  database-backed channels, rather than kernel-managed pipes with
  backpressure.
\end{itemize}

The operating system, having evolved over five decades to solve these
problems for classical software, is less commonly treated as a
first-class runtime substrate for agents. Treating the OS as substrate
also sharpens a deeper question: if the process is the right first
abstraction for agency, where do process semantics stop being expressive
enough for cognition?

\subsection{My Approach}\label{my-approach}

Recent progress in tool calling and structured output makes it practical
to bind natural-language reasoning to a small set of system-level
operations. Instead of introducing another application-layer framework,
I present a runtime architecture where the operating system itself
serves as the execution substrate for agents. The design consists of two
components:

\textbf{Component 1: A Protocol.} A disciplined mapping from agent
concepts to POSIX primitives across four dimensions:

\begin{itemize}
\tightlist
\item
  \textbf{Identity} --- The agent's unique identifier is its process ID
  (PID), assigned by the kernel.
\item
  \textbf{Interface} --- The agent communicates through standard streams
  (\texttt{stdin}/\texttt{stdout}/\texttt{stderr}) and reports outcomes
  via exit status.
\item
  \textbf{State} --- The agent's memory is process memory (cleared on
  exit), environment variables (inherited by children), and filesystem
  (persistent).
\item
  \textbf{Lifecycle} --- The agent spawns children via \texttt{fork},
  continues itself via \texttt{exec}, and terminates via \texttt{exit}.
\end{itemize}

Section 2 elaborates each dimension; Figure 1 illustrates the interface
contract.

\textbf{Component 2: A Single-Image Runtime.} A unitary executable that
implements this protocol. When an agent spawns a child, it instantiates
the same runtime image with different arguments. Parent and child share
code but diverge in state. This recursive structure means the runtime
and the agent template are the same artifact---there is no separate
``agent definition'' language or configuration format. It also means
that delegation is not orchestration from outside: a parent agent
constructs a child's operational world using the same runtime it itself
inhabits.

\subsection{Contributions}\label{contributions}

This paper makes three contributions:

\begin{enumerate}
\def\labelenumi{\arabic{enumi}.}
\item
  \textbf{A systems design perspective for LLM agents.} I argue that the
  operating system can serve as the runtime substrate for agents, and I
  make this claim concrete through a disciplined mapping from agent
  abstractions to POSIX primitives across identity, interface, state,
  and lifecycle (Section 2).
\item
  \textbf{A reference runtime that instantiates this perspective.} I
  present Quine, a single-image runtime in which agents are realized as
  native POSIX processes and recursive delegation is implemented by
  self-instantiation rather than by an external orchestrator (Section
  3).
\item
  \textbf{An analysis of both the reach and the limits of the process
  model for agents.} I show how POSIX process semantics directly yield
  isolation, composition, and self-renewal through exec, and I argue
  that this mapping, precisely because it works, also exposes where
  execution semantics cease to be sufficient for cognition---most
  immediately in the constitution of an agent's world and in the
  revisability of its actions over time (Sections 4--5).
\end{enumerate}

\subsection{Scope and Non-Claims}\label{scope-and-non-claims}

This paper presents the architecture and a reference implementation of
Quine, not a wholesale replacement for existing frameworks. I make no
claims about superior end-task performance; the contribution is a
concrete runtime design point that inherits isolation, composition, and
lifecycle control from POSIX rather than rebuilding them. Its
limitations and boundaries are discussed explicitly in Section 5, which
identifies where process semantics remain effective and where they begin
to leave important aspects of agency unrepresented.

\section{The POSIX Mapping}\label{the-posix-mapping}

Each dimension of the protocol builds on primitives that trace to the
original Unix design (Ritchie and Thompson 1974) and were codified in
(IEEE 2017). The mapping adopts a disciplined one-to-one correspondence:
each agent concept maps to a primary POSIX primitive, and the semantics
are inherited directly from the operating system.

\subsection{Identity}\label{identity}

Identity concerns how an agent instance is distinguished and bounded at
runtime. In Quine, both properties are inherited directly from the
process model.

\begin{itemize}
\tightlist
\item
  \textbf{Agent instance -\textgreater{} Process (PID):} Kernel-managed
  identity.
\item
  \textbf{Agent boundary -\textgreater{} Address space:} Memory
  isolation between agents.
\end{itemize}

The kernel provides a unique identifier---the PID---which is globally
unique within the system for the process lifetime; in the local runtime,
this removes the need for a separate framework-level identifier. This
identifier is used by the scheduler, the memory manager, and the signal
subsystem; Quine inherits these associations rather than reimplementing
them.

The address space boundary defines what memory an agent can access. Two
agents (two processes) cannot read or write each other's memory without
explicit arrangement (shared memory segments, files). This isolation is
enforced by the MMU at hardware level, not by convention or access
control lists in application code.

\subsection{Interface}\label{interface}

Interface concerns how an agent receives instructions, accepts input,
produces output, and signals completion. These interactions map directly
to standard process I/O channels.

\begin{itemize}
\tightlist
\item
  \textbf{Mission (\texttt{argv}):} Immutable mission description.
\item
  \textbf{Material (\texttt{stdin}):} Data input / material.
\item
  \textbf{Deliverable (\texttt{stdout}):} Data output / deliverable.
\item
  \textbf{Diagnostics (\texttt{stderr}):} Diagnostics channel.
\item
  \textbf{Outcome (\texttt{exit\ status}):} Success (0) or failure
  (\textgreater0).
\end{itemize}

The \texttt{argv} is set at process creation and cannot be modified
during execution. This immutability makes it suitable for carrying the
agent's mission---the high-level instruction that defines what the agent
should accomplish. Because \texttt{argv} is separate from
\texttt{stdin}, the instruction channel is structurally distinct from
the data channel.

Standard input (\texttt{stdin}) carries material---the content the agent
operates on---produced by an upstream process or file. Standard output
(\texttt{stdout}) carries the deliverable---the agent's
contribution---consumed by downstream processes or redirection targets.
Standard error (\texttt{stderr}) carries diagnostics---progress
indicators, warnings, and error messages---optionally consumed by
observers without polluting the deliverable stream.

\begin{figure*}[t]
\centering
\includegraphics[width=0.7\textwidth]{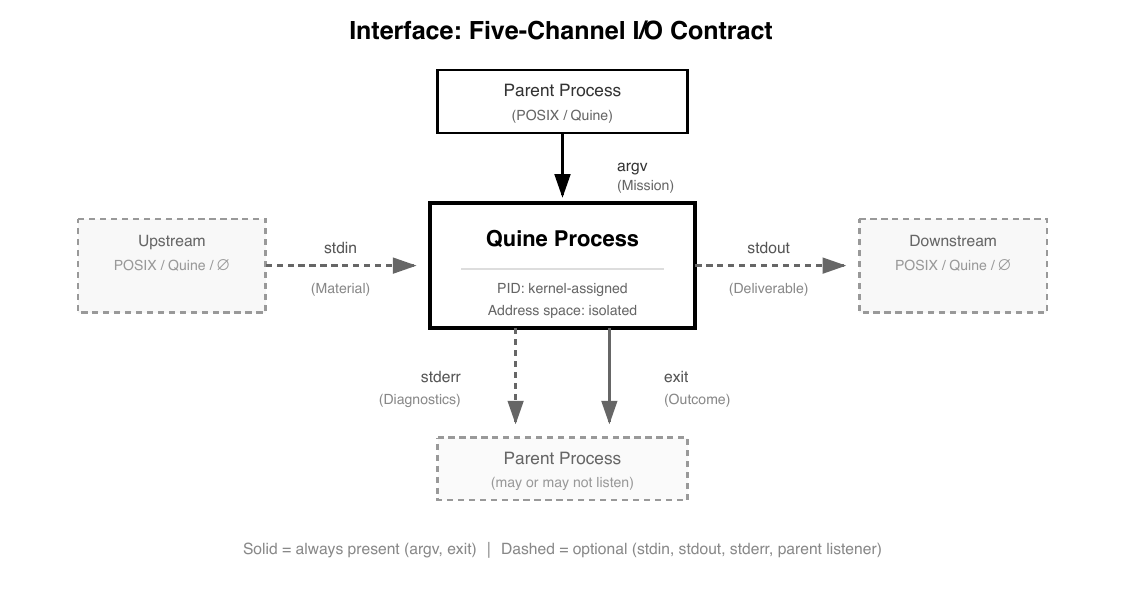}
\caption{The Five-Channel I/O Contract: Mission (argv), Material (stdin), Deliverable (stdout), Diagnostics (stderr), and Outcome (exit status). The parent process provides mission via argv and may listen for diagnostics and outcome. Upstream and downstream processes communicate via standard streams.}
\label{fig:interface}
\end{figure*}

The exit status is a small integer (0--255) that serves as a control
signal indicating the outcome of execution. By convention, 0 indicates
success; non-zero values indicate various failure modes. This signal is
available to the parent process and enables conditional composition in
shell scripts (\texttt{\&\&}, \texttt{\textbar{}\textbar{}}). Figure
\ref{fig:interface} summarizes this five-channel I/O contract.

\subsection{State}\label{state}

State concerns what an agent retains during execution, what survives
across continuation, and what can be externalized for coordination.
Under POSIX, these forms of state map to three tiers with distinct
lifetime scopes.

\begin{itemize}
\tightlist
\item
  \textbf{Ephemeral -\textgreater{} Process memory:} Cleared on
  \texttt{exec}/exit.
\item
  \textbf{Scoped -\textgreater{} Environment variables:} Preserved
  across \texttt{fork} and \texttt{exec}; inherited by children as
  independent copies.
\item
  \textbf{Global -\textgreater{} Filesystem:} Persistent; shared
  (default) or isolated via namespaces.
\end{itemize}

\textbf{Ephemeral state} (process memory) exists only while the process
runs. When the process calls \texttt{exec} or terminates, this state is
lost. For LLM agents, this corresponds to the ``working memory''
accumulated during a single execution---the context window contents,
intermediate computations, and any in-memory data structures.

\textbf{Scoped state} (environment variables) survives both
\texttt{fork} and \texttt{exec}. On \texttt{fork}, the child inherits a
copy; on \texttt{exec}, the calling process retains its environment.
This makes environment variables suitable for passing compact metadata
between generations: progress markers, configuration flags, or
compressed summaries of prior execution. The copy-on-fork semantics mean
children can modify their environment without affecting the parent.

\textbf{Global state} (filesystem) persists beyond process lifetime and
is visible to all processes (subject to permissions). This is the only
state tier that survives both \texttt{exec} and process termination. For
agents, the filesystem serves as long-term memory, shared artifacts, and
coordination medium.

\subsection{Lifecycle}\label{lifecycle}

Lifecycle concerns how agents are created, delegate work, renew
themselves, and terminate. These transitions are expressed directly
through process control primitives.

\begin{itemize}
\tightlist
\item
  \textbf{Spawn (\texttt{fork}):} Create child process with new mission.
\item
  \textbf{Continue (\texttt{exec}):} Replace current process image while
  preserving process-level continuity.
\item
  \textbf{Terminate (\texttt{exit(status)}):} Signal outcome to parent.
\end{itemize}

An agent can \textbf{spawn} children to delegate work (synchronously via
\texttt{wait}, or asynchronously via background execution and job
control). Each child receives its own \texttt{argv} (a distinct
sub-mission) and inherits environment variables and context history from
the parent. The parent can block until the child completes, or continue
executing while monitoring child status through signals and job control.

Beyond structural delegation, \texttt{fork} provides cognitive
decomposition: each child operates with an independent context window,
allowing complex problems to be partitioned into subproblems that
individually fit within context limits. The parent aggregates results
without carrying the full reasoning burden of each subtask.

An agent can \textbf{continue} itself by calling \texttt{exec} with its
own image. This replaces the process image---clearing process
memory---while preserving PID, parent relationship, environment
variables, and (optionally) open file descriptors. In the general POSIX
case, \texttt{exec} may also install a different executable image and a
different \texttt{argv}. Quine's default self re-entry path reuses its
current image and mission \texttt{argv}, so the agent can reset
cognitive context while maintaining the same directive. For LLM agents,
this provides a mechanism to escape context limits: the agent can
checkpoint progress in environment variables (or offload to the
filesystem for larger state), then \texttt{exec} to start fresh with a
clean context window while continuing the same task.

\begin{figure*}[t]
\centering
\includegraphics[width=\textwidth]{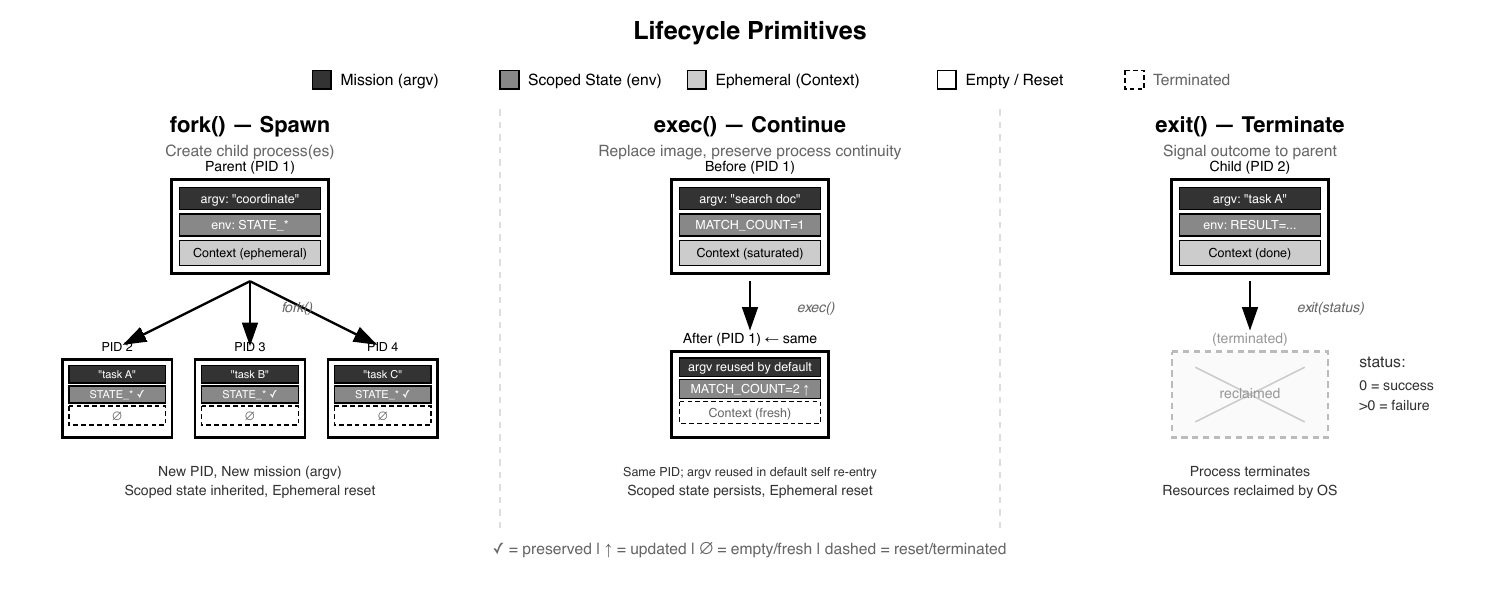}
\caption{Agent Lifecycle: The three lifecycle operations (spawn, continue, terminate) and their relationship to state tiers. Spawn creates child processes with distinct missions; continue replaces the process image while preserving process-level continuity, with mission continuity arising in Quine's default self re-entry path; terminate signals completion to the parent.}
\label{fig:lifecycle}
\end{figure*}

An agent \textbf{terminates} by calling \texttt{exit} with a status
code. This releases all resources (memory, file descriptors, child
processes if the parent does not wait), notifies the parent, and makes
the exit status available for inspection. The three operations---spawn,
continue, terminate---combined with the state hierarchy, define the
lifecycle model illustrated in Figure \ref{fig:lifecycle}.

The next section describes how this mapping is realized in a concrete
runtime.

\section{The Runtime}\label{the-runtime}

The implementation is a unitary executable that serves as both runtime
and agent template.

\subsection{Host-Guest Architecture: Separation of Control and
Compute}\label{host-guest-architecture-separation-of-control-and-compute}

\begin{figure*}[t]
\centering
\includegraphics[width=0.5\textwidth]{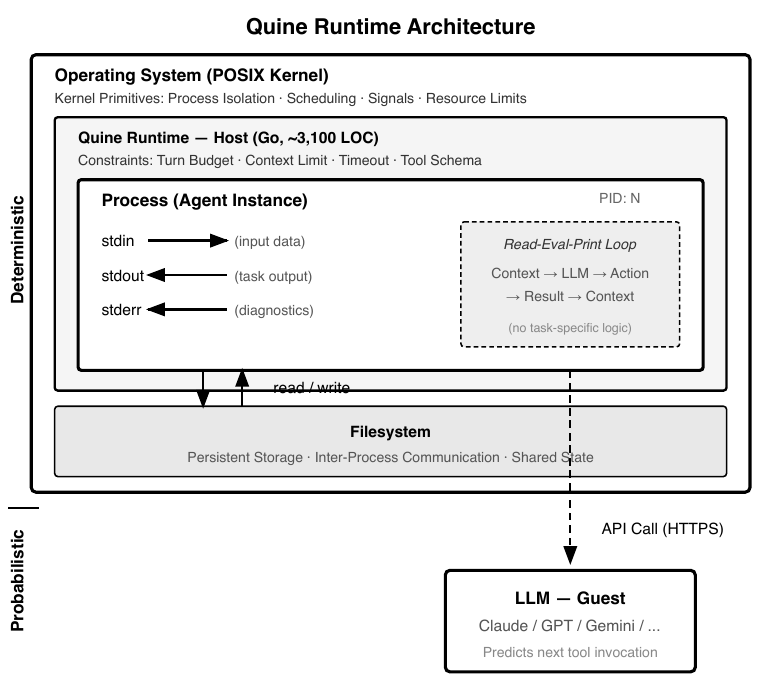}
\caption{Host-Guest Architecture: The runtime separates deterministic control (Host) from probabilistic computation (Guest). The Host manages syscalls, file descriptors, and signals; the Guest provides decisions, tool calls, and reasoning.}
\label{fig:architecture}
\end{figure*}

Quine enforces a physical separation between deterministic control flow
and probabilistic computation, as shown in Figure
\ref{fig:architecture}. This separation is not merely architectural
preference; it reflects a fundamental asymmetry in where state lives and
how failures manifest.

\textbf{Host (Local OS Process).} The Host is a conventional compiled
program that maintains the agent's context, lifecycle, and filesystem
state. It is responsible for:

\begin{itemize}
\tightlist
\item
  Parsing \texttt{argv} to extract the mission
\item
  Providing an annotated shell environment with file descriptors
  (including stdin) for the Guest
\item
  Serializing context (conversation history, tool results) for
  transmission to the Guest
\item
  Managing child processes spawned by the Guest's tool invocations
\item
  Calling \texttt{exit} with the status code determined by the Guest
\end{itemize}

The Host maintains conversation history, control state, and I/O buffers
for tool execution. Because the Host is an ordinary OS process, agent
instances inherit standard resource management: they can be scheduled,
signaled, killed, or resource-limited by POSIX tools (\texttt{nice},
\texttt{kill}, \texttt{ulimit}, \texttt{cgroups}) without requiring
framework-specific supervision.

\textbf{Guest (Remote LLM API).} The Guest acts as a stateless cognitive
oracle. It receives the serialized POSIX state (mapped to prompts) and
returns structured tool invocations. The Guest has no persistent state
between calls; all context must be provided in each request. This
statelessness is a design choice: it ensures that the Host maintains
authoritative state and that the Guest can be replaced, rate-limited, or
load-balanced without coordination.

The Host contains no task-specific logic; decision-making is delegated
to the Guest. The Host is purely reactive: it provides the environment
the Guest operates in, captures the results, and iterates.

\textbf{Context Assembly.} The Host preserves OS-level I/O boundaries
when assembling context:

\begin{enumerate}
\def\labelenumi{\arabic{enumi}.}
\tightlist
\item
  The mission (\texttt{argv}) is placed in the System Prompt.
\item
  Material (\texttt{stdin}) is announced in the User Message; access is
  provided via file descriptor remapping (Section 3.2).
\item
  Tool outputs are returned as Tool Result messages.
\end{enumerate}

This layered structure aligns with instruction hierarchy training
(Wallace et al. 2024), where models prioritize System \textgreater{}
User \textgreater{} Tool. The Host does not rewrite or summarize
streams; it annotates boundaries and delegates interpretation to the
Guest.

\textbf{Tool Interface.} The Host exposes four tools---\texttt{sh},
\texttt{fork}, \texttt{exec}, \texttt{exit}---named after their POSIX
counterparts with semantics as defined in Section 2.4:

\begin{itemize}
\tightlist
\item
  \textbf{\texttt{sh}:} Execute shell command, return stdout/stderr/exit
  status.
\item
  \textbf{\texttt{fork}:} Spawn child Quine process(es) with new
  \texttt{argv}; optionally block until completion.
\item
  \textbf{\texttt{exec}:} Replace current process image; preserve pid,
  env vars, and file descriptors. By default Quine re-execs its own
  binary with the current mission \texttt{argv}, but explicit
  target/\texttt{argv} can hand control to a different executable.
\item
  \textbf{\texttt{exit}:} Terminate with status code.
\end{itemize}

The \texttt{sh} tool is the primary interaction mechanism---file
operations, compilation, and system commands all go through shell
invocations. The \texttt{fork} tool enables delegation without leaving
the process model; each child receives its own mission and returns
structured results. The \texttt{exec} tool is how agents manage context
growth: in the default self re-entry path, agents checkpoint progress in
env vars or the filesystem, then \texttt{exec} into a fresh Quine
instance with the same mission and a clean context window. The same
primitive can also hand off directly to a non-Quine executable when the
task is better finished as an ordinary POSIX process.

\subsection{POSIX Conformance: File Descriptor
Mapping}\label{posix-conformance-file-descriptor-mapping}

From the shell's perspective, Quine is a standard POSIX filter: stdin
in, stdout out, stderr for diagnostics, exit status for outcome. This
means Quine composes anywhere a traditional Unix filter can: pipelines,
shell scripts, subprocesses.

The challenge is maintaining clean context windows while honoring this
contract. The runtime solves this by exposing annotated file descriptors
within the shell environment. The runtime's stdin, stdout, and stderr
are passed into each shell invocation as higher-numbered file
descriptors:

\begin{itemize}
\tightlist
\item
  \textbf{fd 3:} Runtime's stdin (the material stream)
\item
  \textbf{fd 4:} Runtime's stdout (the deliverable channel)
\item
  \textbf{fd 5:} Runtime's stderr (the failure-signal channel)
\end{itemize}

This mapping leaves the shell's standard file descriptors (0, 1, 2)
available for normal command I/O. To read input material, the Guest
reads from fd 3. To emit a deliverable, the Guest writes to fd 4 (e.g.,
\texttt{echo\ "result"\ \textgreater{}\&4}). To emit streaming failure
signals, the Guest writes to fd 5 (e.g.,
\texttt{echo\ "failed"\ \textgreater{}\&5}). The shell command's own
stdout/stderr are captured and returned to the Guest as tool results for
reasoning.

This separation allows the Guest to capture command output for reasoning
while emitting deliverables to downstream processes through a separate
channel. Without it, the agent would face a dilemma: pollute the
deliverable stream with intermediate outputs, or lose visibility into
command results.

\subsection{Illustrative Shell
Compositions}\label{illustrative-shell-compositions}

Quine's adherence to standard streams enables composition with classic
Unix utilities and multi-agent pipelines.

\textbf{Example 1: Single-process replacement.} A cognitive filter that
understands intent:

\begin{Shaded}
\begin{Highlighting}[]
\FunctionTok{cat}\NormalTok{ server.log }\KeywordTok{|} \ExtensionTok{./quine} \StringTok{"Extract lines indicating auth failures"} \KeywordTok{|} \FunctionTok{sort} \KeywordTok{|} \FunctionTok{uniq} \AttributeTok{{-}c}
\end{Highlighting}
\end{Shaded}

Here Quine replaces \texttt{grep} in a traditional pipeline. The
cognitive filter receives log lines on stdin, applies semantic
understanding to identify authentication failures, and emits matching
lines to stdout. Downstream tools (\texttt{sort}, \texttt{uniq\ -c})
process the output normally.

\textbf{Example 2: Implicit DAG.} Three agents form a reasoning chain:

\begin{Shaded}
\begin{Highlighting}[]
\FunctionTok{git}\NormalTok{ diff HEAD\textasciitilde{}1 }\KeywordTok{|} \DataTypeTok{\textbackslash{}}
  \ExtensionTok{./quine} \StringTok{"Identify the changed components"} \KeywordTok{|} \DataTypeTok{\textbackslash{}}
  \ExtensionTok{./quine} \StringTok{"Assess risk level for each change"} \KeywordTok{|} \DataTypeTok{\textbackslash{}}
  \ExtensionTok{./quine} \StringTok{"Generate a review checklist"}
\end{Highlighting}
\end{Shaded}

Each agent receives the previous agent's output as stdin and contributes
its analysis to stdout. The shell provides the topology; no orchestrator
or shared memory is required. Each agent runs in a separate process with
separate context.

\textbf{Example 3: Control flow without framework.} Exit codes drive
branching:

\begin{Shaded}
\begin{Highlighting}[]
\ExtensionTok{./quine} \StringTok{"Apply the patch"} \OperatorTok{\textless{}}\NormalTok{ fix.patch }\KeywordTok{\&\&} \DataTypeTok{\textbackslash{}}
  \ExtensionTok{./quine} \StringTok{"Run the test suite and report results"}
\end{Highlighting}
\end{Shaded}

The first agent attempts to apply a patch; it exits 0 on success,
non-zero on failure (e.g., conflicts, malformed patch). The shell's
\texttt{\&\&} operator conditionally executes the second agent only if
the patch applied cleanly. This is cognitive branching using standard
shell control flow.

Because Quine is packaged as a standard POSIX executable, the same
composability extends reflexively: a Quine instance may invoke
\texttt{./quine\ "mission"} through the \texttt{sh} tool. This should be
understood not as a second lifecycle primitive, but as an external
composition path enabled by the single-image design. The runtime's
internal delegation mechanism remains \texttt{fork}; shell-mediated
re-invocation is a consequence of POSIX conformance, not a separate
runtime protocol.

\subsection{Implementation Scale}\label{implementation-scale}

The Host is implemented in approximately 3,100 lines of Go (cognitive
loop, tool execution, job management, and session recording). LLM
provider abstractions add another 2,600 lines (protocol adapters, OAuth
flows, and configuration). The complete implementation totals
approximately 5,700 lines, compiled to a single \textasciitilde9.8 MB
binary that supports multiple LLM providers (Anthropic Claude, OpenAI
GPT, Google Gemini) through a provider abstraction layer. Provider
selection is controlled by environment variables, enabling the same
binary to use different backends without recompilation.

The runtime contains no domain-specific task logic; decision-making is
delegated to the Guest (LLM). The properties that emerge from this
design---containment, composition, and continuity---are examined in the
next section.

\section{Properties}\label{properties}

The previous sections described the mapping and the runtime; this
section examines what the design yields. The properties discussed
here---Containment, Composition, and Continuity---are not implemented
features; they emerge from representing agents as POSIX processes. By
grounding agents in the process model, Quine inherits mechanisms refined
over five decades of Unix evolution.

\subsection{Containment}\label{containment}

Containment in Quine is inherited from the layered enforcement structure
of the POSIX process model, not added by the runtime. Each layer
constrains a different dimension of agent behavior, and higher layers
cannot override lower-layer guarantees.

\textbf{Hardware and kernel enforcement.} Each agent occupies a distinct
address space enforced by the MMU; a child crash cannot corrupt the
parent's memory. The kernel further constrains resource consumption
(cgroups, ulimit, OOM killer) and privilege (capabilities, namespaces,
seccomp). When limits are exceeded or unauthorized operations attempted,
enforcement occurs at the kernel level---the process is terminated with
an observable cause. These bounds apply to agents that are buggy or
adversarial: fork bombs, unauthorized network access, and container
escapes are bounded by kernel enforcement, not application-layer
exception handlers.

\textbf{Runtime: instruction-data separation.} Above these layers, Quine
preserves a boundary at the interface level: mission (argv) and material
(stdin) travel through distinct OS channels. The Host maintains this
separation when constructing the LLM context: argv maps to the System
Prompt, stdin content enters only the User Message. Combined with
instruction hierarchy training (Wallace et al. 2024), this yields a
structural basis that may reduce some prompt manipulation
risks---control over input data does not by itself grant control over
the instruction channel. This is a claim about architectural separation,
not a complete security guarantee.

The result is a containment model that is inherited rather than
reimplemented: hardware isolates memory, the kernel bounds resources and
privileges, and the runtime preserves a structural distinction between
instruction and data.

\subsection{Composition}\label{composition}

Composition arises from two sources: recursive delegation through
\texttt{fork}, and external invocability as a standard POSIX command.
The former is a runtime mechanism; the latter is a structural
consequence of the process model.

\textbf{Recursive composition via \texttt{fork}.} A child process is
structurally isomorphic to its parent: same executable, same interface
(argv/stdin/stdout/stderr/exit status). This removes the distinction
between ``manager'' and ``worker''---any instance can delegate by
spawning children, and any child can coordinate its own subtree. From
the parent's perspective, a delegated subtree remains encapsulated
behind the process boundary.

This recursive structure also distributes cognitive load. Each child
begins with an independent context window and reasons only about its
subproblem; the parent carries only the child's returned result, not its
intermediate reasoning. Problems exceeding a single agent's reasoning
budget can be partitioned into tractable subproblems, with the process
tree serving as an implicit divide-and-conquer structure.

\textbf{Shell-level composability.} Because Quine conforms to the
standard command contract, it is directly invocable by the shell as an
ordinary POSIX executable. The shell can launch, sequence, redirect, and
supervise Quine exactly as it would any other Unix command---composing
with pipelines, \texttt{xargs}, GNU \texttt{parallel}, \texttt{cron}, or
existing CI workflows without adaptation. No separate workflow language,
daemon, or application-layer protocol is required.

This composability extends reflexively: a Quine instance may invoke
\texttt{./quine\ "sub-mission"} through the \texttt{sh} tool, treating
another agent as an ordinary command. This is not a second delegation
primitive but a consequence of POSIX conformance---the runtime's
internal mechanism remains \texttt{fork}; shell-mediated invocation is
simply the external view of the same single-image design.

\subsection{Continuity}\label{continuity}

Agents face two forms of mortality: context exhaustion (cognitive death)
and process termination (physical death). Quine uses standard POSIX
mechanisms to persist across both.

\textbf{Surviving cognitive death: continuation across \texttt{exec}.}
The \texttt{exec} syscall replaces the process image---clearing memory
and conversational context---while preserving process ID, parent
relationship, environment variables, and open file descriptors. In
Quine's default self re-entry path, the current image and mission
\texttt{argv} are reused as well. This gives the agent a way to renew
itself without becoming a different computational entity in the process
graph.

As an agent approaches context limits, it can \texttt{exec} into a fresh
instance while preserving several distinct continuity surfaces at once:
process identity in the process graph, live material and pipeline state
in open file descriptors, and scoped or durable state in environment
variables or the filesystem. In Quine's default self re-entry path,
mission continuity is preserved as well through reuse of the current
\texttt{argv}. The key idea is not to preserve raw cognition, but to
preserve enough structured state to make cognition reconstructible.
\texttt{wisdom} is one Quine-specific convenience for encoding compact
scoped state in environment variables, not the definition of
\texttt{exec} continuity itself. Long-lived history can remain on the
filesystem; active context is selectively reassembled after renewal.

\textbf{Surviving physical death: feedback through stderr and exit
status.} When a child process terminates---whether by completing its
task or failing---the parent must decide how to continue. A failing
child emits diagnostics on stderr before terminating; the parent reads
this and decides whether to retry, skip, or escalate. Exit status
(0--255) provides a compact outcome signal; shell conditionals
(\texttt{\&\&}, \texttt{\textbar{}\textbar{}}) become cognitive branch
points. Standard Unix supervision semantics thus function as adaptive
agent coordination.

\subsection{Operational Validation}\label{operational-validation}

The following observations demonstrate that the architecture is
operational---agents can exercise these properties in practice. These
are qualitative feasibility demonstrations, not performance benchmarks.
Detailed execution traces are provided in Appendix A.

\textbf{Composition: recursive delegation.} In an exploratory search
task exceeding single-agent budgets, agents used \texttt{fork} to spawn
parallel workers, assigned disjoint sectors, and coordinated results
through the filesystem. One run produced a 3-level process tree (36
sessions) mirroring the target directory structure---demonstrating
recursive delegation and inter-process coordination. (Appendix A.1)

\textbf{Continuity: exec-based self-renewal.} In MRCR-style needle
retrieval tasks (OpenAI 2024), agents processed material ranging from 4K
to 279K tokens via stdin. Short contexts required no renewal; long
contexts triggered adaptive \texttt{exec} cycles---one run required 9
cycles, externalizing progress to environment variables between each
renewal. A baseline comparison (loading full context without streaming)
failed on 5 of 8 samples; the streaming architecture with \texttt{exec}
renewal succeeded on all. (Appendix A.2)

\textbf{Reproducibility.} Implementation, prompts, and execution logs
are available at https://github.com/kehao95/quine.

The same mapping that yields these properties also clarifies where Quine
sits among existing systems and where the POSIX model stops being
sufficient.

\section{Related Work and the Boundaries of
POSIX}\label{related-work-and-the-boundaries-of-posix}

Quine sits among existing agent systems as a runtime-level alternative;
from there, the discussion turns to where the POSIX model ends and what
must extend beyond it.

\subsection{Related Systems}\label{related-systems}

I classify current agent systems by their structural relationship to the
operating system. The taxonomy is organized by where agent identity,
lifecycle, interface, and isolation are realized---not by end-task
capability or developer-facing features.

\subsubsection{Host-Coupled Extensions}\label{host-coupled-extensions}

Systems such as Cursor and GitHub Copilot are integrated within host
environments such as IDEs and do not provide agents with independent
process lifecycles. The agent component exists within the host
application rather than as a separately managed runtime entity: it is
not exposed as a first-class POSIX process, does not present an
independent stdin/stdout interface, and terminates with the host
application. This tight coupling enables deep integration---such as
inline suggestions and direct access to editor state---but
correspondingly limits independent spawning, termination, and
shell-level composition with external tools.

\subsubsection{Application-Layer
Schedulers}\label{application-layer-schedulers}

Frameworks such as LangGraph, AutoGen (Wu et al. 2023), and CrewAI
(CrewAI, Inc. 2024) provide user-space scheduling and message buses.
Agents are realized as in-process objects (Python classes, coroutines)
with fault isolation implemented through exception handling rather than
hardware-enforced address space separation. These systems have
successfully demonstrated multi-agent coordination and have large
ecosystems of tools and integrations; their design optimizes for
developer productivity and rapid prototyping. Structurally, agents
within these frameworks typically share a single OS process: a crash in
one agent (unhandled exception, memory corruption) can propagate to
others, and scheduling is managed by the framework dispatcher rather
than the kernel.

\subsubsection{Sandboxed Loops}\label{sandboxed-loops}

Systems such as Devin, OpenHands (Wang et al. 2024), and SWE-agent (Yang
et al. 2024) use an external controller to manage containerized tools.
The agent's ``body'' (shells, browsers, file access) runs in an isolated
sandbox; the ``brain'' (LLM) runs in the controller process. This
architecture provides strong tool isolation---a runaway shell command
cannot escape the container---while centralizing cognitive coordination,
and has proven effective for complex software engineering tasks.
Structurally, however, the sandbox isolates tools rather than agents
themselves: the cognitive loop (prompt -\textgreater{} LLM
-\textgreater{} action -\textgreater{} observation) runs in a single
controller process, and multiple ``agents'' are typically coroutines or
threads within this controller rather than separate OS processes.

\subsubsection{Pseudo-OS Middleware}\label{pseudo-os-middleware}

Systems such as AIOS (Mei et al. 2024), agentOS (Li et al. 2026), and
UFO 2 ({Zhang et al.} 2025) implement OS-like abstractions in user
space, providing schedulers, memory managers, and IPC mechanisms for
agents using terminology borrowed from operating systems. These systems
recognize that agent management resembles process management and attempt
to provide similar abstractions. However, despite OS-inspired naming,
agents in these systems are typically objects or threads within a single
host process. ``Process isolation'' is simulated through software
boundaries rather than hardware-enforced address spaces; the
``scheduler'' is a user-space dispatcher rather than the kernel's CFS;
resources are accounted at the framework level rather than by cgroups or
ulimit.

\subsubsection{Quine's Position}\label{quines-position}

Quine represents a distinct design point. Existing agent runtimes
typically either manage agents at the application layer or use the OS
primarily as a tool sandbox, rather than directly realizing each agent
as a native POSIX process with the full reasoning-and-acting loop
exposed through standard process interfaces.

This difference is about runtime organization, not end-task superiority.
Application-layer frameworks offer flexibility, rich ecosystems, and
rapid development cycles. Quine trades some of this flexibility for
structural properties inherited from the OS: kernel scheduling rather
than framework dispatch, standard streams rather than framework-specific
message formats, process lifecycle primitives rather than object
instantiation, and hardware-enforced isolation rather than exception
handling.

Having located Quine among current systems, I now turn from comparative
structure to the limits of the substrate itself. The POSIX mapping
provides a valid first abstraction for agents, but process semantics do
not exhaust the runtime needs of cognition. Among the directions that
boundary exposes, two are especially immediate: one concerns
space---whether an agent's world can be scoped by relevance rather than
only permission; the other concerns time---whether cognition and
side-effects can be revised on different terms. Both arise directly from
taking the process model seriously: once the agent is granted an
execution boundary, the next question is what world that boundary
encloses, and what kind of temporality its actions inhabit.

\subsection{From Namespace to World}\label{from-namespace-to-world}

Plan 9 showed that a process need not inhabit a single global
filesystem: namespaces can be per-process, and interfaces can be
constructed rather than merely inherited (Pike et al. 1993). For agents,
this insight is necessary but insufficient. An agent requires not merely
a different namespace, but a world organized by task relevance---a
situated perspective in which some entities are present, others absent,
and still others foregrounded or merely nameable.

A subjective world is not a false world; it is a selectively constituted
one. The distinction matters because it separates runtime responsibility
from epistemic illusion. When a debugging agent sees logs, stack traces,
and failing tests while a planning agent sees deadlines, owners, and
design intents, the difference is not merely what files each may access.
It is what objects are present as first-class entities at all. Security
asks what an agent may access; subjectivity asks what its world is made
of.

This reframing has architectural consequences. Traditional sandboxing
restricts resources; a cognitive runtime must do more---it must scope
reality. The agent does not merely execute within constraints; it is
situated within a world whose boundaries are drawn by relevance, role,
and task. Two agents on the same machine, with access to the same
underlying storage, may nonetheless inhabit different operational worlds
if their runtimes foreground different entities and relations. In such a
runtime, the difference may appear not only in which paths are mounted,
but in which objects are surfaced as logs, goals, owners, hypotheses, or
pending obligations.

In Quine, this constitutive role does not belong to an external control
plane. Because the runtime is recursively self-instantiating, the same
executable that inhabits a world can also construct a different world
for a child agent. A parent does not merely launch a subprocess; it
defines the visible environment into which that subprocess is born. The
agent is therefore neither a passive resident of a pre-given environment
nor the object of a separate orchestrator's worldview---it is a
constitutive participant in the production of local worlds, for itself
through renewal and for others through delegation. The distinctive point
is not only that worlds can be scoped, but that world-construction is
endogenous to the runtime itself.

POSIX can isolate processes; Plan 9 can differentiate namespaces. But a
cognitive runtime may need to expose worlds whose constitution is
task-relative rather than permission-derived. This does not yet define a
mechanism; it marks a shift in what the runtime must be responsible for
making visible.

\subsection{From Execution to
Revision}\label{from-execution-to-revision}

POSIX time is operational and forward-moving: actions happen, effects
accumulate, processes terminate. Rollback, where available, is external,
partial, and typically resource-centric---restoring a file, replaying a
log, restarting a container. But cognition is not merely sequential; it
is provisional. An agent may reconsider, backtrack, explore
alternatives, and retain lessons from failed attempts.

The core difficulty is that conventional rollback conflates two kinds of
state that agents need to treat differently. Mental state comprises
beliefs, plans, branches explored, rejected hypotheses, and learned
constraints. Environmental state comprises files changed, commands
executed, resources allocated, and messages sent. When these cannot be
revised independently, the choice is stark: either both are lost, or
neither is reversible. What agents need is rollback without
amnesia---the ability to undo effects while preserving experience. A
failed branch may need to retract file edits and spawned processes while
preserving the constraints it discovered and the options it ruled out.

This is not simply an engineering problem of better checkpointing. Agent
work often involves speculative branches: multiple candidate futures
explored from a common past, some committed, others abandoned. A runtime
that only knows committed execution cannot treat branching as
first-class; it reduces counterfactual exploration to ad hoc application
logic, with each framework inventing its own replay mechanism.

POSIX manages process lifetime but does not express the provisionality
of cognition. If revisability is central to how agents think, then time
itself becomes part of the runtime contract---not a library feature
bolted on afterward, but a structural concern that the operating layer
must acknowledge. The question is not how to implement snapshots, but
whether alternative futures can be explored without reducing them to
workarounds above the OS.

\subsection{Note on Scope}\label{note-on-scope}

The two directions above---world and time---do not exhaust the
boundaries of the POSIX model. Internal cognitive structure (making the
agent's reasoning addressable rather than opaque) and distributed
composition (preserving file-based abstractions across machine
boundaries) mark additional frontiers. This paper focuses on the POSIX
mapping itself; these extensions belong to future work.

The broader question---what an operating system for cognition should
expose, delimit, remember, and compose---is closer to the Plan 9 lineage
(Pike et al. 1995) than to conventional POSIX extension. The aim is not
merely to add mechanisms, but to rethink what the runtime should make
visible, nameable, and composable.

\section{Conclusion}\label{conclusion}

Quine demonstrates that the operating system can serve as a first-class
runtime substrate for LLM agents, not merely an execution sandbox for
their tools. By mapping agent identity, interface, state, and lifecycle
to POSIX process semantics, this architecture replaces application-layer
orchestration with kernel primitives.

The model requires accepting Unix assumptions: deterministic composition
through pipes, text-stream interfaces, and shared-nothing isolation.
These prove to be productive constraints. Delegating isolation,
scheduling, and resource control to the OS yields containment enforced
from hardware, composition via recursive delegation and shell utilities,
and self-renewal across context limits through \texttt{exec}.

This mapping also exposes where process semantics become insufficient
for cognition. The architectural mismatches identified---unrepresented
internal structure, undifferentiated worldviews, irreversible time, and
local-bound composition---mark boundaries for future work. Modern
kernels have absorbed much of the Plan 9 lineage; the question is how to
compose these primitives at the runtime layer.

Source code: \href{https://github.com/kehao95/quine}{repository}

\protect\phantomsection\label{refs}

\appendix

\section{Appendix A: Qualitative Evidence for System
Properties}\label{appendix-a-qualitative-evidence-for-system-properties}

\small

This appendix provides detailed qualitative evidence for the properties
discussed in Section 4. These observations come from exploratory runs;
they demonstrate operational feasibility, not statistical claims.

\subsection{Composition: Recursive
Delegation}\label{composition-recursive-delegation}

Composition in Quine arises from the \texttt{fork} primitive and
standard shell mechanisms. When agents face tasks that exceed their
individual execution budget, they can delegate subtasks to children via
\texttt{fork}, coordinating through the filesystem or stdout.

\subsubsection{Observation: Fractal Library
Search}\label{observation-fractal-library-search}

In an exploratory run, an agent was tasked with searching 1,000 files
for anomalous content. The agent had 8 turns---insufficient for
sequential inspection of all files. The agent used the available
primitives as follows:

\begin{enumerate}
\def\labelenumi{\arabic{enumi}.}
\tightlist
\item
  \textbf{Explored structure} (Turn 1): Used \texttt{ls} to identify the
  library's hex/shelf/volume hierarchy.
\item
  \textbf{Spawned children} (Turn 2): Forked 10 parallel workers at
  depth 1, each assigned a disjoint hex sector.
\item
  \textbf{Recursive delegation}: Some children themselves forked,
  creating 25 workers at depth 2. Total: 36 sessions across 3 levels.
\end{enumerate}

\textbf{Delegation intent example:}

\begin{Shaded}
\begin{Highlighting}[]
\ExtensionTok{[CONTEXT]:}\NormalTok{ Search the library directory ./library/hex\_01/}
           \ControlFlowTok{for}\NormalTok{ something }\ExtensionTok{non{-}random.}
\ExtensionTok{[GOAL]:}\NormalTok{ Find which file }\ErrorTok{(}\ControlFlowTok{if} \ExtensionTok{any}\KeywordTok{)} \ExtensionTok{contains}\NormalTok{ non{-}random content.}
\ExtensionTok{[DELIVERABLE]:}\NormalTok{ If found, write the full filepath to stdout.}
\ExtensionTok{[VERIFY]:}\NormalTok{ cat the file you identify and confirm}
          \ExtensionTok{it}\NormalTok{ contains non{-}random content.}
\end{Highlighting}
\end{Shaded}

The resulting structure was self-similar: each level mirrored the
library's own hex/shelf/volume hierarchy. This demonstrates that the
\texttt{fork} primitive supports recursive decomposition and that agents
can coordinate through structured delegation.

\subsection{Continuity: Exec-Based
Self-Renewal}\label{continuity-exec-based-self-renewal}

When an agent's context window approaches exhaustion, it can use
\texttt{exec} to replace its process image with a fresh instance while
preserving process-level continuity, live file descriptors, and any
externalized state carried through environment variables or filesystem
artifacts. In the MRCR runs discussed here, Quine used its default self
re-entry path, so mission continuity was preserved as well.

\subsubsection{Observation: MRCR Long-Context
Retrieval}\label{observation-mrcr-long-context-retrieval}

In MRCR-style needle retrieval tasks (OpenAI 2024), agents received:

\begin{itemize}
\tightlist
\item
  \textbf{Mission (argv):} ``Find the sixth short essay about distance;
  prepend the hash and output.''
\item
  \textbf{Material (stdin):} 4K--279K tokens of streaming conversation
  data
\end{itemize}

The architecture enforces a structural separation between mission (argv)
and material (stdin): argv content enters the System Prompt, stdin
content enters only the User Message. This separation remained stable
throughout all runs---material content never entered the instruction
channel.

\textbf{Adaptive self-renewal.} Short contexts (4K--7K tokens) required
no \texttt{exec} calls; the agent completed tasks in single sessions.
Long contexts triggered adaptive renewal cycles:

\begin{itemize}
\tightlist
\item
  178K tokens: 9 \texttt{exec} cycles, \textasciitilde50 read operations
\item
  279K tokens: \textasciitilde12 \texttt{exec} cycles, \textasciitilde80
  read operations
\end{itemize}

Across these renewals, \texttt{exec} provided several continuity
channels at once: - Cleared the conversation context (cognitive renewal)
- Preserved mission and process identity for the continuing run -
Preserved the stdin stream position (material continuity) - Preserved
open stdio for downstream completion and tool re-entry - Allowed compact
progress state to be carried via \texttt{wisdom} / environment variables
when useful, with larger artifacts left on the filesystem

\textbf{One compact state-transfer example:}

\begin{Shaded}
\begin{Highlighting}[]
\FunctionTok{\{}
  \DataTypeTok{"found\_count"}\FunctionTok{:} \StringTok{"4"}\FunctionTok{,}
  \DataTypeTok{"current\_position"}\FunctionTok{:} \StringTok{"\textasciitilde{}100K tokens"}\FunctionTok{,}
  \DataTypeTok{"partial\_content"}\FunctionTok{:} \StringTok{"..."}
\FunctionTok{\}}
\end{Highlighting}
\end{Shaded}

\textbf{Baseline comparison.} When the same tasks were attempted by
loading the full conversation into a single LLM context (without
argv/stdin separation or streaming), the model failed on 5 of 8 samples:

\begin{itemize}
\tightlist
\item
  Short contexts (4K--7K, 5 samples): Quine \(\geq 0.996\) all; Baseline
  failed 3/5
\item
  Long contexts (178K--279K, 3 samples): Quine 1.000 all; Baseline
  failed 3/3
\end{itemize}

This demonstrates that the \texttt{exec} primitive enables reliable
processing of inputs that exceed single-context capacity, and that the
architecture supports but does not require self-renewal.

\subsection{Cross-Property
Interaction}\label{cross-property-interaction}

These properties can operate together in practice. In the fractal
library search (A.1), some child agents used \texttt{exec} with wisdom
when approaching their turn limits:

\begin{Shaded}
\begin{Highlighting}[]
\FunctionTok{\{}
  \DataTypeTok{"children\_pids"}\FunctionTok{:} \StringTok{"48991, 48996, 48998, 49008, 49016"}\FunctionTok{,}
  \DataTypeTok{"content\_observed"}\FunctionTok{:} \StringTok{"All files contain UUID{-}like strings..."}\FunctionTok{,}
  \DataTypeTok{"files\_checked"}\FunctionTok{:} \StringTok{"volume\_00000.txt, volume\_00001.txt, ..."}\FunctionTok{,}
  \DataTypeTok{"next\_action"}\FunctionTok{:} \StringTok{"Check child results for non{-}random findings"}\FunctionTok{,}
  \DataTypeTok{"strategy"}\FunctionTok{:} \StringTok{"Spawned 5 children to search hex\_00, 01, 02, 05, 09"}
\FunctionTok{\}}
\end{Highlighting}
\end{Shaded}

This demonstrates that the architecture supports Composition (fork for
parallelization) and Continuity (exec for self-renewal) operating
together within the Containment boundary (each process isolated, unable
to corrupt siblings).

\subsection{Reproducibility}\label{reproducibility}

Implementation, prompts, and execution logs for these observations are
available in the \href{https://github.com/kehao95/quine}{repository}.
The MRCR experiments use samples from the OpenAI MRCR benchmark (OpenAI
2024).


\begin{CSLReferences}{1}{1}
\bibitem[\citeproctext]{ref-CrewAI}
CrewAI, Inc. 2024. \emph{CrewAI: Framework for Orchestrating
Role-Playing Autonomous AI Agents}. GitHub repository.

\bibitem[\citeproctext]{ref-IEEEPOSIX}
IEEE. 2017. \emph{Standard for Information Technology---Portable
Operating System Interface (POSIX)}. IEEE Std 1003.1-2017. IEEE.

\bibitem[\citeproctext]{ref-LangChain}
LangChain, Inc. 2023. \emph{LangChain: Building Applications with LLMs
Through Composability}.
\href{https://github.com/langchain-ai/langchain}{Https://github.com/langchain-ai/langchain}.

\bibitem[\citeproctext]{ref-Li2026}
Li, Chen, Xiaoyu Liu, Xiang Meng, and Xin Zhao. 2026. {``Architecting
AgentOS: From Token-Level Context to Emergent System-Level
Intelligence.''} \emph{arXiv Preprint arXiv:2602.20934}.

\bibitem[\citeproctext]{ref-McIlroy1978}
McIlroy, M. D. 1978. {``Unix Time-Sharing System: Foreword.''}
\emph{Bell System Technical Journal} 57 (6): 1899--904.

\bibitem[\citeproctext]{ref-Mei2024}
Mei, Kai, Zelong Li, Shuyuan Xu, Ruosong Ye, Yingqiang Ge, and Yongfeng
Zhang. 2024. {``AIOS: LLM Agent Operating System.''} \emph{arXiv
Preprint arXiv:2403.16971}.

\bibitem[\citeproctext]{ref-OpenAIMRCR}
OpenAI. 2024. \emph{Multi-Turn Retrieval Context Reasoning (MRCR)
Benchmark}.
\href{https://huggingface.co/datasets/openai/mrcr}{Https://huggingface.co/datasets/openai/mrcr}.

\bibitem[\citeproctext]{ref-Pike1995}
Pike, Rob, Dave Presotto, Sean Dorward, et al. 1995. {``Plan 9 from Bell
Labs.''} \emph{Computing Systems} 8 (3): 221--54.

\bibitem[\citeproctext]{ref-Pike1993}
Pike, Rob, Dave Presotto, Ken Thompson, and Howard Trickey. 1993. {``The
Use of Name Spaces in Plan 9.''} \emph{Operating Systems Review} 27 (2):
72--76.

\bibitem[\citeproctext]{ref-RitchieThompson1974}
Ritchie, Dennis M., and Ken Thompson. 1974. {``The UNIX Time-Sharing
System.''} \emph{Communications of the ACM} 17 (7): 365--75.

\bibitem[\citeproctext]{ref-Wallace2024}
Wallace, Eric, Kai Xiao, Reimar Leike, Lilian Weng, Johannes Heidecke,
and Alex Beutel. 2024. {``The Instruction Hierarchy: Training LLMs to
Prioritize Privileged Instructions.''} \emph{arXiv Preprint
arXiv:2404.13208}.

\bibitem[\citeproctext]{ref-Wang2024}
Wang, Xingyao, Boxuan Chen, Ziniu Adelt, et al. 2024. {``OpenHands: An
Open Platform for AI Software Developers as Generalist Agents.''}
\emph{arXiv Preprint arXiv:2407.16741}.

\bibitem[\citeproctext]{ref-Wu2023}
Wu, Qingyun, Gagan Bansal, Jieyu Zhang, et al. 2023. {``AutoGen:
Enabling Next-Gen LLM Applications via Multi-Agent Conversation
Framework.''} \emph{arXiv Preprint arXiv:2308.08155}.

\bibitem[\citeproctext]{ref-Yang2024}
Yang, John, Carlos E. Jimenez, Alexander Wettig, Kilian Liber, Karthik
Narasimhan, and Ofir Press. 2024. {``SWE-Agent: Agent-Computer
Interfaces Enable Automated Software Engineering.''} \emph{arXiv
Preprint arXiv:2405.15793}.

\bibitem[\citeproctext]{ref-Zhang2025}
{Zhang, Chaoyun, He Huang, Chao Ni, et al.} 2025. {``UFO2: The Desktop
AgentOS.''} \emph{arXiv Preprint arXiv:2504.14603}.

\end{CSLReferences}
\end{document}